\documentclass[conference]{IEEEtran}
\usepackage[nocompress]{cite}
\usepackage[utf8]{inputenc}
\usepackage{graphicx}
\usepackage{amsmath}
\usepackage{amsfonts}
\usepackage{amssymb}
\usepackage{algorithm}
\usepackage{algorithmic}
\graphicspath{{images/}{../images/}}
\usepackage{subfiles}
\usepackage{fancyhdr}
\usepackage{xcolor}
\usepackage{multirow}
\usepackage{textcomp}
\usepackage{array}
\usepackage{tabu}
\usepackage{amsmath}
\usepackage{bm}
\usepackage{caption}
\usepackage{subcaption}

\usepackage{calc}
\usepackage{lipsum}
\usepackage{hyperref}

\def \fullmethod {COded Taking And Giving}
\def \method {COTAG}
\def \link {l}
\def \labelunsnt {L'}
\def \labelsnt {k}
\def \labelrecv {L_{\link}}

\newcommand{\newadd}[1]{{\color{black}{#1}}}
\newcommand{\response}[1]{{\color{black}{#1}}}

\newcommand{\ICC}[1]{{\color{black}{#1}}}
\newcommand{\ICCdel}[1]{{\color{black}{#1}}}
\newcommand{\ICCadd}[1]{{\color{black}{#1}}}

\newcommand{\ICCfinal}[1]{{\color{black}{#1}}}

\title{COded Taking And Giving (COTAG): Enhancing Transport Layer Performance over \ICC{Indoor} Millimeter Wave Access Networks
}

\author{\IEEEauthorblockN{Zongshen~Wu\IEEEauthorrefmark{1},
Chin-Ya~Huang\IEEEauthorrefmark{2}, \sublargesize\normalfont\textit{Member, IEEE}, and
Parameswaran~Ramanathan\IEEEauthorrefmark{1}, \sublargesize\normalfont\textit{Fellow, IEEE}
}
\IEEEauthorblockA{\IEEEauthorrefmark{1}Department of Electrical and Computer Engineering,
University of Wisconsin, Madison, WI, 53706 USA
}
\IEEEauthorblockA{\IEEEauthorrefmark{2}Department of Electronic and Computer Engineering,\\ 
National Taiwan University of Science and Technology, Taipei, 106, Taiwan\\
}
\IEEEauthorblockA{E-mail: zongshen.wu@wisc.edu,
chinya@mail.ntust.edu.tw, parmesh.ramanathan@wisc.edu}
\vspace{-16pt}
}

\begin{document}

\IEEEtitleabstractindextext{%

\begin{abstract}
Millimeter wave (mmWave) access networks have the potential to meet the high-throughput and low-latency needs of immersive applications. However, due to the highly directional nature of the mmWave beams and their susceptibility to beam misalignment and blockage resulting from user movements and rotations, the associated mmWave links are vulnerable to large channel fluctuations. These fluctuations result in  disproportionately adverse effects on performance of transport layer protocols such as Transmission Control Protocol (TCP). 
To overcome this challenge, we  propose a network layer solution, \textit{\textbf{\fullmethod\ (\method)}} scheme to sustain low-latency and high-throughput end-to-end TCP performance \ICCadd{in dually connected networks}. In particular, \method\ creates network encoded packets at the network gateway and each access point (AP) aiming to adaptively take the spare bandwidth on  each link for transmission. Further, if one link bandwidth drops due to user movements, \method\ actively abandons the transmission opportunity by conditionally dropping packets. Consequently, \method\ actively \ICC{adapts to link quality changes in mmWave access network} and enhances the TCP performance without jeopardizing the latency of immersive content delivery. To evaluate the effectiveness of the proposed \method, we conduct experiments using off-the-shelf APs and network simulations. The evaluation results show that \method\ improves end-to-end TCP performance significantly on both throughput and latency.
\end{abstract}
\begin{IEEEkeywords}
mmWave Links, TCP, \ICCfinal{Dual Connectivity}, Immersive Content, Network Coding.
\end{IEEEkeywords}}

\maketitle
\thispagestyle{plain}
\pagestyle{plain}
\IEEEdisplaynontitleabstractindextext
\IEEEpeerreviewmaketitle

\section{Introduction}
The demand for immersive content such as augmented
reality (AR) and virtual reality (VR) is growing rapidly. These
applications require high-throughput
and low-latency network connectivity to
support their rich, dynamic, and interactive content. They also
need wireless connectivity for untethered access. In this paper,
we assume that the wireless connectivity is \ICC{primarily} provided using
millimeter wave (mmWave) based communication link.
Note that, mmWave links have the potential to support high throughput
because there is a large amount of spectrum
available in this frequency band, e.g., Federal Communications
Commission (FCC) has allocated 7 GHz (57 GHz – 64 GHz)
of contiguous unlicensed spectrum for commercial and public
use.

As the mmWave band brings the channel capacity to a new
high level, there are also challenges. One challenge with the
mmWave band is that the beams are substantially more directional
than the conventional sub-6 GHz band. Furthermore,
link bandwidth degrades more severely at the mmWave band
than at the sub-6 GHz band \cite{Sun2018, TS}. For example, the 60 GHz signal
attenuates 15 to 20 dB for the first-order reflected path as
compared to the line-of-sight (LOS) path. Although adaptive
beam management can alleviate deterioration in channel conditions,
detecting a change in the channel conditions as well
as reforming the beams takes time. The delays incurred in
adapting the beams adversely affect the end-to-end latency and
the resulting end-to-end throughput, which in turn degrades
AR/VR experience. \ICCfinal{Additionally}, at certain frequencies especially
around the 60 GHz band, the mmWave channels are highly
susceptible to human blockages.
\ICCfinal{Finally, although dual connectivity reduces end-to-end outage probability\cite{WirelessURLLC}, the link quality imbalance in dual connectivity can impact upper-layer end-to-end throughput and latency.}

\ICC{
AR/VR applications often use Transmission Control Protocol (TCP) to ensure reliable, sequenced delivery of packets from server to client. Due to the above characteristics of mmWave band \ICCadd{and due to changes in user pose, location, and direction during AR/VR applications}, TCP’s performance often falls well short of its performance in a comparable fully wired network. This is because human-induced channel quality deteriorations often causes the data rate to fall substantially. As a result, the bits in transit from the TCP sender to the TCP receiver experience severe congestion, which is sensed by the TCP sender after a certain network delay. Even after the lower layer protocols have largely restored the data rates, the TCP sender cannot sense this increase. Instead, in order to preserve network fairness, it slowly strives to increase the sending rate. While the sending rate is slowly \ICCfinal{increasing}, the end-to-end throughput is substantially below the available bandwidth.

Although there are several potential solutions to this problem, it is very difficult to avoid the inherent delay for solutions at the transport layer to work effectively. 
\ICCfinal{The motivation of this paper is to enhance the transport layer end-to-end performance for indoor applications like AR/VR in dual connectivity mmWave network by alleviating impacts of channel fluctuations and link quality imbalance.} 
We propose a network layer technique called \textbf{\textit{\fullmethod\ (\method)}} 
\ICCfinal{in this paper}. \method\ assumes that there is a dual wireless connectivity to end user, through either two mmWave links or one mmWave link and one sub-6G link. An advantage of dual connectivity is that there will be no complete loss of connectivity unless both links are blocked simultaneously. \method\ encodes packets into linear combinations at the gateway of indoor wireless network and forwards these combinations of packets through the dual connectivity for decoding at the mobile device. During the forwarding, if spare bandwidth is available, \method\ also generates extra combinations to fully use the bandwidth. \method\ adaptively gives up transmission opportunity when one link does not have enough bandwidth and generates redundancy when there is adequate bandwidth. In this way, \method\ adapts to the changes in link quality and alleviates adverse effects of channel fluctuations and link quality imbalance to enhance TCP performance.
}

The rest of this paper is organized as follows. 
The related work is introduced in Section~\ref{sec:releated}. 
The proposed approach and algorithms are described in Section~\ref{sec:proposed}. 
The evaluation results demonstrating the effectiveness of the proposed mechanism are presented in Section~\ref{sec:eval}. 
Finally, the paper is summarized in Section~\ref{sec:summary}.
 


\section{Related Work}
 
\label{sec:releated}
\noindent
{\bf MPTCP.}
Transmission Control Protocol (TCP) is the most common transport layer scheme for reliable, sequenced
delivery of data. 
There are many variants of TCP.
The recent variants such as NewReno \cite{newreno}, CUBIC \cite{Ha2008}, Compound \cite{ctcp}, and BBR \cite{BBR} are designed for networks
with high bandwidth delay products (BDP). However, they are not well suited for dealing with large
bandwidth changes due to fluctuations in the wireless channel quality. 
\ICCfinal{Moreover, although TCP variants for wireless networks have been studied for dealing with packet errors and single channel fluctuations \cite{wirelesstcp}, further enhancement is needed for mmWave based dual connectivity networks to mitigate impacts from not only packet loss, but also imbalance in link quality.}
The TCP variant most related
to the solution in this paper is Multipath TCP (MPTCP) \cite{mptcp}. 

MPTCP is a transport layer protocol designed to better utilize the bandwidth 
available across multiple paths in order to increase the end-to-end throughput. 
It tends to work well when the multiple paths from the source to the receiver are
mostly edge disjoint. If there are many shared links among the paths, then the effectiveness is
diminished.
The effectiveness of MPTCP has been evaluated in delivering across multiple interfaces of an end host
in \cite{Polese2017}. \cite{Polese2017} shows that the effectiveness of
MPTCP is not satisfactory when the bandwidth of mmWave link fluctuates significantly over time.

\response{
\ICCadd{To alleviate this problem,}
MuSher \cite{musher} adjusts the packet forwarding ratio 
according to the transport layer throughput and buffer conditions \ICCadd{across the multiple paths in MPTCP}.
The adaptive packet scheduling at the transport layer in Musher enhances MPTCP performance in dual connectivity network with mmWave \ICCadd{(IEEE 802.11ad)} and sub-6G \ICCadd{(IEEE 802.11ac)} links. 
\ICCadd{As compared} to Musher, this paper addresses a network layer enhancement to the dual connectivity network for AR/VR scenario, where the frequent user movements, rotations and blockages result in significant channel fluctuations. 


}

\noindent
{\bf SRNC.} Spare-bandwidth Rate-adaptive Network Coding (SRNC) is an approach
to enhance TCP performances in a wireless mesh based network with gigabit links.
The scheme proposed in this paper is inspired by the concept of network coding in SRNC.
Due to the mesh networks, there are many paths with possibly many links with available bandwidth to
provide many opportunities for network coding in \cite{SRNC}. Since packets from TCP flows destined for different
end hosts are cross-coded, the decoding also occurs in the wireless mesh network. The end host is
delivered unencoded packets and it is completely oblivious to the network coding that occurred in the mesh.
\ICCadd{Such cross-coding is not applicable to scenarios considered in this paper.}
 
\noindent
{\bf Other Related Work.}
Challenges of mmWave communications have been addressed at all layers of network stack. At
physical layer, Sur et al. propose MUST to predict the best beam and to redirect user traffic over
conventional WiFi when there is blockage in the mmWave link \cite{Sur2017}. Although MUST focuses on the
physical layer, it needs support from higher layers of the protocol to redirect
user traffic.\newadd{\cite{WirelessURLLC} discusses the challenges and tradeoffs of Ultra-Reliable Low Latency Communication (URLLC) considering massive MIMO and multi-connectivity.}
Using mmWave links cooperatively with LTE, 5G new radio (NR) are also being investigated in \cite{Polese2017, Yang2018}.
\ICCfinal{Moreover, relaying video with mmWave based data plane and sub-6G based control plane is proposed in \cite{spider} to mitigate impacts of link failures in mmWave based relay networks.}
\ICCfinal{Non-Orthogonal Multiple Access (NOMA) supports the multi-user wireless networks for 5G NR in both power and code domains \cite{nomasurvey}. Besides the improvement achieved by NOMA for multi-user cases, enhancement at the higher layer is also important for more robust dual connectivity which may be supported by different base stations connected with the same user equipment.}


\section{ \fullmethod\ (\method) Scheme}
\label{sec:proposed}

\subsection{Overview}

\begin{figure}
\begin{subfigure}[b]{0.5\columnwidth}
\centering
  \includegraphics[width=\columnwidth]{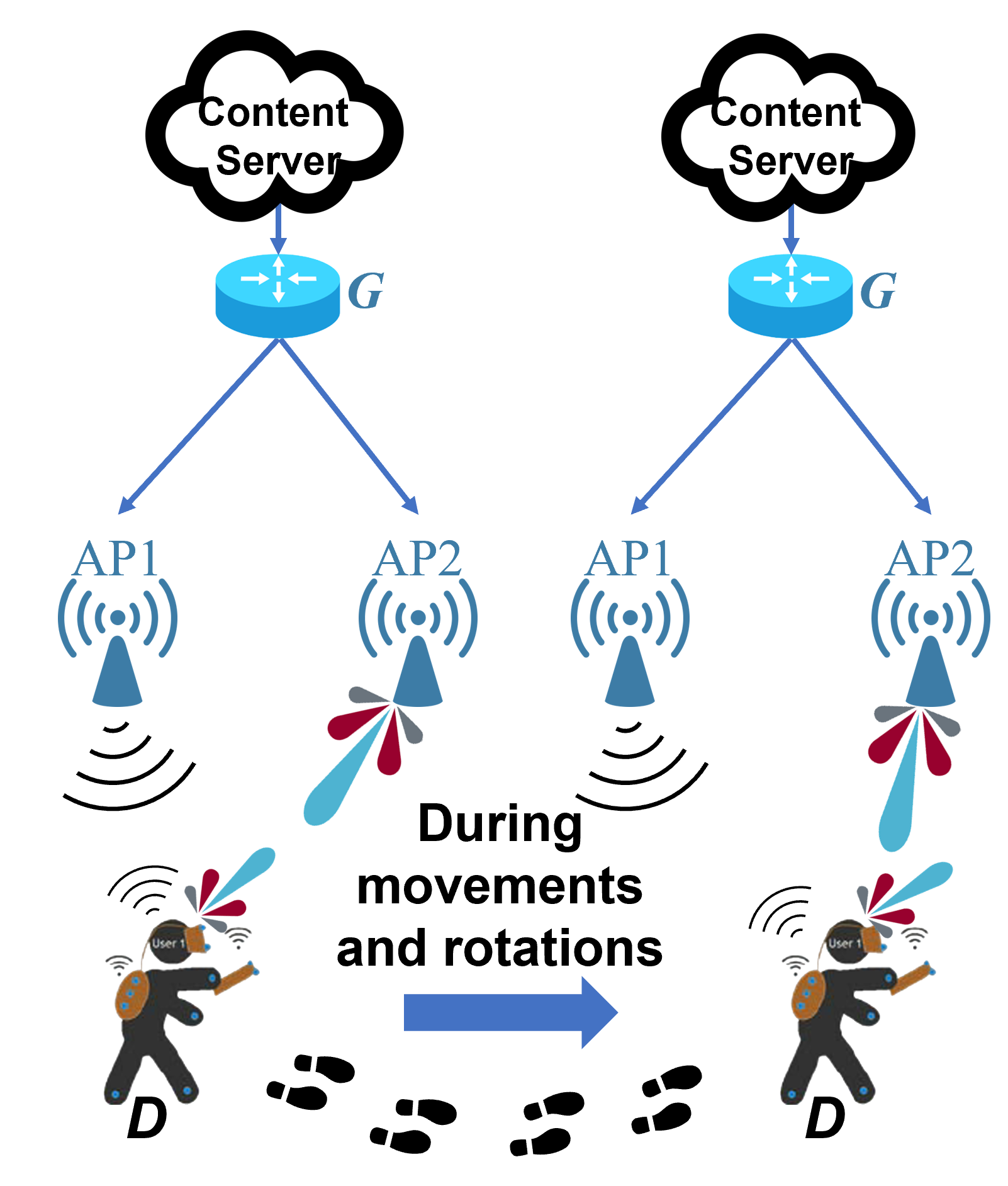}
\caption{Illustration of a network with sub-6G and mmWave access} 
\label{fig:simple_topo}
\end{subfigure}
\hspace{10pt}
\begin{subfigure}[b]{0.4\columnwidth}
\centering
  \includegraphics[width=\columnwidth]{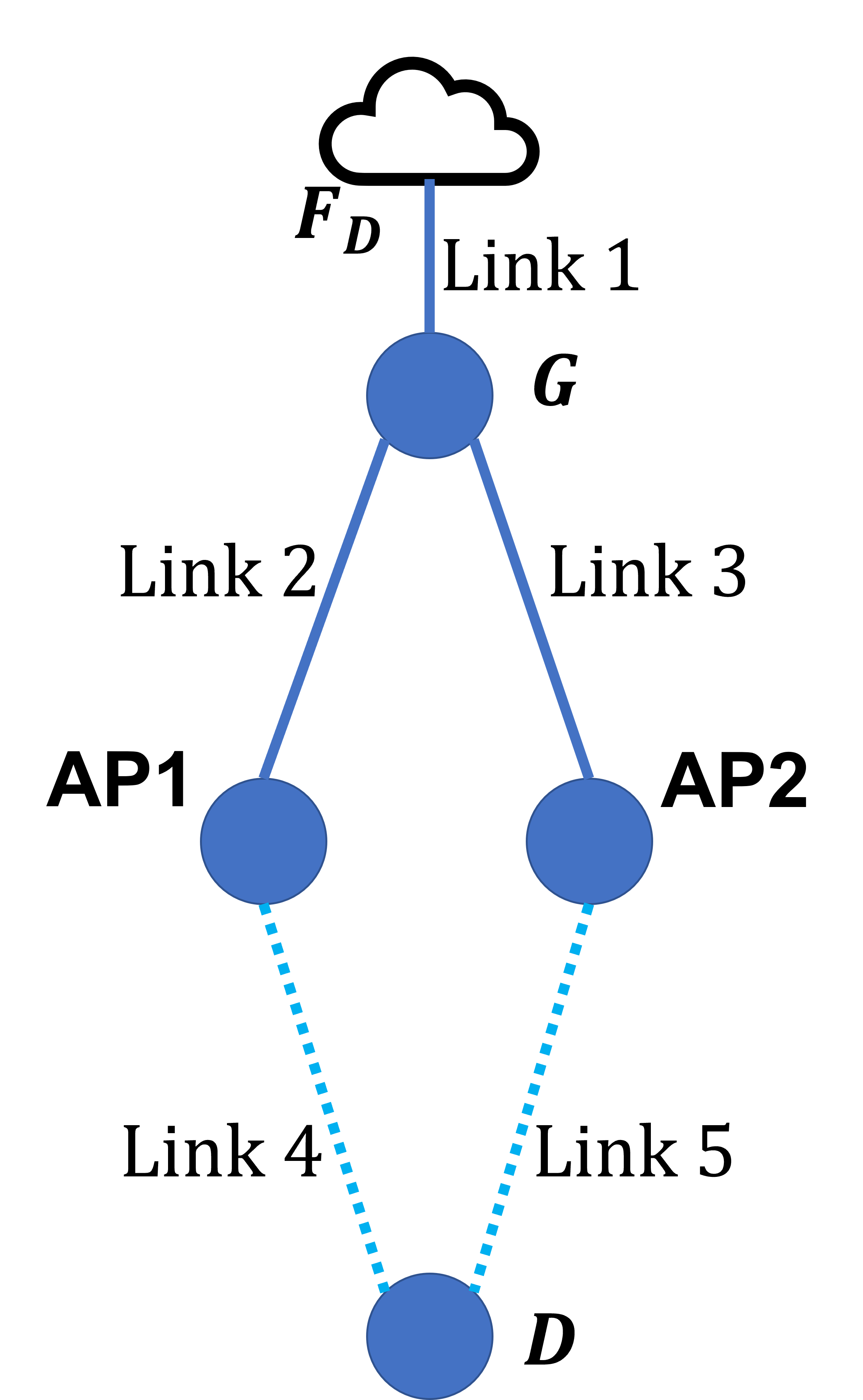}
\caption{Simple topology} 
\label{fig:vtopo}
\end{subfigure}
\caption{AR/VR dual connectivity network topology} 
\vspace{-15pt}
\end{figure}

For simplicity of presentation, consider a simple topology
formed by an immersive content server: a network gateway $G$, \ICCadd{one sub-6G AP (AP1), one mmWave
AP (AP2),} and a mobile device $D$ (e.g., AR/VR headset) (see 
Fig.~\ref{fig:simple_topo} and~\ref{fig:vtopo}). Assume that the mobile device moves in a small geographic area in the vicinity of these two APs as a
part of the immersive experience. \response{In the immersive experience, the user moves from one place to another. During this process, the user induces changes in pose, location and direction, which result in mmWave channel quality variations.}

Gateway $G$ is on the route from the content server to the device $D$. It forwards the packets belonging to the traffic flows $F_D$ that are associated with the delivery of the immersive content.
\ICCdel{Specifically, the traffic flows $F_D$ may be comprised of multiple TCP flows, multiple flows can help achieve higher throughput because it is easier to deal with slower growth in the congestion window of each flow after packet losses or reordering effects.} 
Aiming to provide low-latency and \ICC{high} TCP
throughput to the mobile device, when the network gateway $G$ receives
packets of $F_D$, it forwards them to the two APs. Device $D$ is simultaneously connected to both APs. It may receive packets
from both APs at the same time. \method\ is a network layer solution
\ICCadd{that complements the} different strategies implemented at the physical and link layers.
The key aspect of the proposed scheme is that the forwarding of the packets at $G$ and the two APs \newadd{does} not occur in the traditional fashion. Instead, they forward using concepts of \ICC{{\em coded forwarding}, {\em coded taking}, and {\em giving}}  as described below.

\begin{figure*}[!ht]
    \centering
    \includegraphics[width=0.99\textwidth]{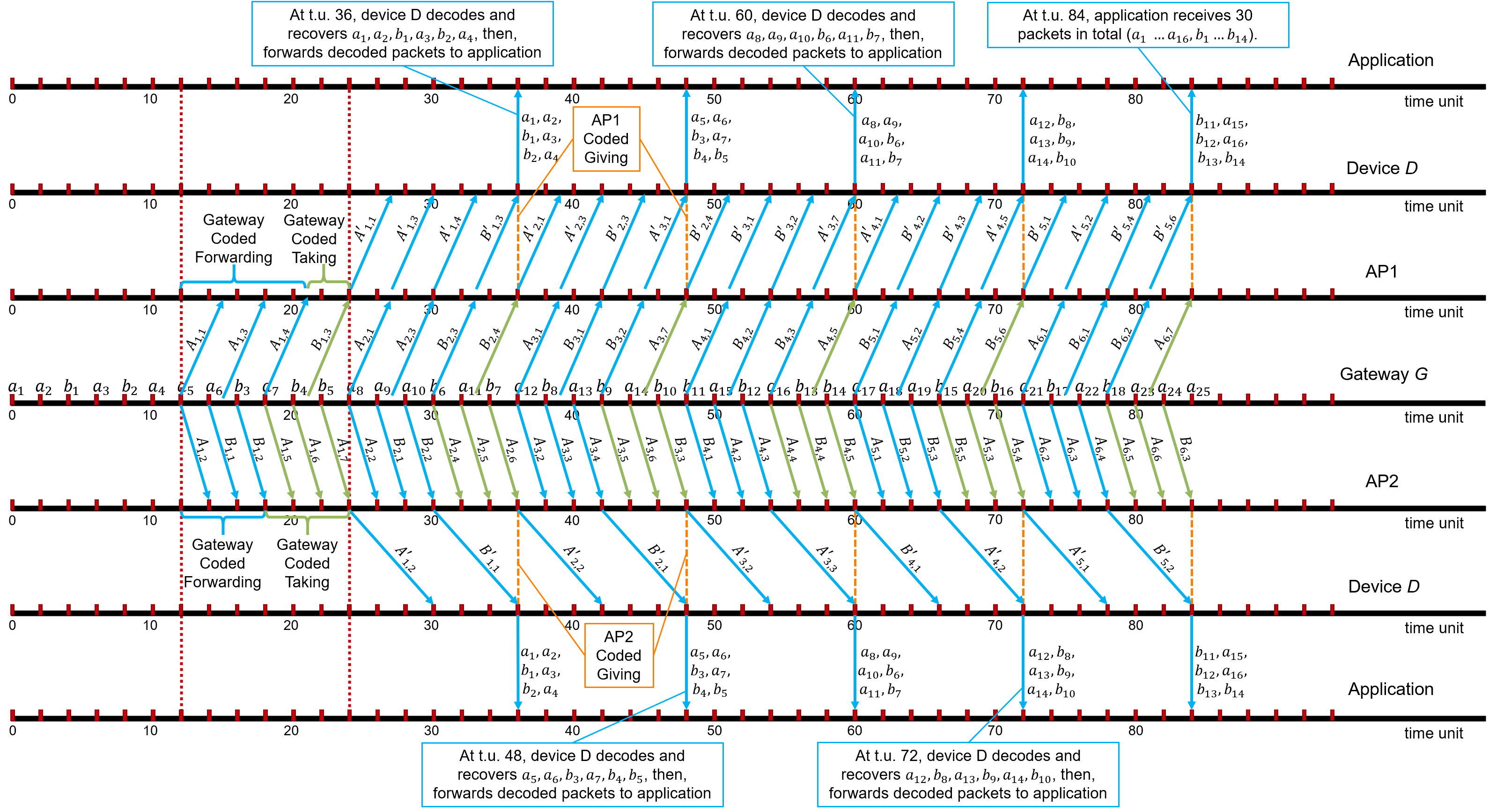}
    \caption{Illustration of cases with \method\ \label{fig:COTAGtu}}
\end{figure*}


\subsection{\method\ Illustration}
Fig.~\ref{fig:COTAGtu} shows how sequenced packets at gateway $G$ are delivered to the application using \method. For simplicity of explanation, this example assumes that $G$, AP1, AP2, and $D$ are tightly synchronized. In reality, as detailed in later sections of this paper, the \method\ is self-clocking and synchronization between the nodes is not needed. \method\ introduces three \ICC{concepts: {\em Coded Forwarding}, {\em Coded Taking}, and {\em Giving}}.

\textbf{Coded Forwarding:} Instead of forwarding packets in its queue, each node ($G$, AP1, and AP2) forwards “network coded” packets, which are linear combinations of a cohort of packets from the same TCP flow. To create a cohort of packets, \method\ a priori selects a “buffering time” denoted by $T$. In Fig.~\ref{fig:COTAGtu}, the value of $T$ is 12. All packets arriving at $G$ in the interval $[(k-1)T, kT)$, $k$ = 1, 2, …, are buffered until time $kT$. These packets form cohort “$k$”.  In Fig.~\ref{fig:COTAGtu}, all packets that arrive at $G$ in the interval [0, 12) form cohort 1. They are not forwarded until time 12. Similarly, packets which arrive at $G$ in the interval [12, 24) form cohort 2, and so on. At time 12, $G$ starts forwarding packets from cohort 1. The first packet in the queue from cohort 1 is $a_1$. Therefore, $G$ linearly combines all cohort 1 packets from the “$a$” flow ($a_1$, $a_2$, $a_3$, and $a_4$) to generate a network coded packet $A_{1,1}$ and forwards it to AP1. The second packet in the queue from cohort 1 is $a_2$. Therefore, $G$ once again generates a network coded packet $A_{1,2}$ by combining $a_1$, $a_2$, $a_3$, and $a_4$ and forwards it to AP2. After $A_{1,2}$ is transmitted, the third packet in the queue is $b_1$. Therefore, $G$ linearly combines $b_1$ and $b_2$ and forwards it to AP2. Gateway $G$ proceeds similarly until all packets in the queue have been considered or it is time for the next cohort.
Similarly,
nodes AP1 and AP2 also buffer incoming packets and forward linearly coded packets from the same cohort. 

\textbf{Coded Taking:} $G$, AP1, and AP2 are not done with forwarding packets in a cohort after they have been through the buffer once during the time interval. For example, at time 18, $G$ is done considering all the six packets in cohort 1 once. However, the time allocated for transmitting cohort 1 does not end till time 24. Between time 18 and 24, $G$ forwards redundant packets. Specifically, it reconsiders the first, second, third, and so on packets in cohort 1, generates appropriate network coded packets and forwards them until either the time expires or a specified redundancy limit is achieved. Since $a_1$ is the first packet in this cohort, at time 18, $G$ forwards $A_{1,5}$, which is a linear combination of $a_1$, $a_2$, $a_3$, and $a_4$. In this example, for cohort 1, $A_{1,5}$, $A_{1,6}$, $A_{1,7}$, and $B_{1,3}$ are redundant packets; there are six packets in cohort 1 but $G$ forwarded a total of 10 packets for cohort 1 to AP1 and AP2.
AP1 and AP2 also follow the same strategy, although the example does not show this situation.

\textbf{Giving:} Sometimes, depending on the link bandwidth and the number of packets in the cohort, $G$, AP1, and AP2 may not have adequate time to even consider all packets in the cohort at least once. In this case, “Giving” is invoked. Unlike in conventional forwarding, $G$, AP1, and AP2 does not forward all packets that are queued. If the time for a cohort runs out before all the packets are forwarded, the nodes simply discard the remaining packets and transition to handling the next cohort. For example, AP2 receives six packets of cohort 1 in the interval [12, 24). However, due to low data rate of the mmWave link between AP2 and $D$, AP2 is only able to forward two of these packets, in a network coded fashion, in the interval [24, 36). At time 36, AP2 discards all cohort 1 packets and starts forwarding packets from cohort 2.

\subsection{\method\ Mechanism}
Specifically, the proposed scheme \method\ has a parameter $T$ that represents the scheme interval. The \method\ scheme applies interval parameter $T$ to arrange arrived packets in cohorts for \ICC{Coded Forwarding, Coded Taking and Giving.} All packets of $F_D$ that arrive at $G$ in
the time interval $[(k-1)T, kT)$, are queued and forwarded
\underline{only} in the interval $[kT, (k+1)T)$, for integer $k \geq 1$. Let us refer to the packets of $F_D$ which arrive in the time interval $[(k-1)T, kT)$, as cohort $k$ packets.

\begin{algorithm}[!h]
\small
{\bf In time interval} $[kT, (k + 1)T)$:
\begin{algorithmic}[1]
\IF{$t$ == $kT$,}
\STATE [Coding: ] Discard all packets in $F_{D}$ which arrive in $[(k-2)T, (k-1)T)$.
\STATE [Coding: ]  With $n_{k}$ packets arrived in $[(k-1)T, kT)$, create $n_{k}$ random linear network encoded packets in $F_{D}$ arriving in $[(k-1)T, kT)$.
\STATE [Coding: ]  Attach a label $k$ to each encoded packet.
\STATE [Coding, Give: ]  Mark these network encoded packets as ``High priority''.
\ENDIF

\STATE [Coding: ] Buffer all incoming packets in $F_{D}$.

\hspace{-18pt}\noindent\textbf{\newadd{On each outgoing link:}}\\
\hspace{-12pt}\noindent Transmit (as permitted by
          the packet scheduler) network \\
\hspace{-12pt}\noindent encoded packets :
\IF{the total number of transmitted packets of all outgoing links is less than $n_{k}$,}
\STATE [Coding: ]  Forward the  ``High priority'' network encoded packets.
\ELSE
\STATE [Take: ]   Generate a random linear combined packet of packets in $F_{A}$.
\STATE Attach a label $k$, mark it as ``Low priority'', and forward it.
\ENDIF

\end{algorithmic}
\caption{{Algorithm executed by the network \textbf{Gateway} in \textbf{\method}.}}
\label{alg:gateway}
\end{algorithm}

\noindent
{\textbf{Operations at} \bf Gateway \bm{$G$}:} Forwarding at $G$ in the interval $[kT, (k+1)T)$ occurs in the following way with the description in Algorithm.~\ref{alg:gateway}.
\begin{itemize}
\item {\bf Coded Forwarding.} All cohort $k$ packets  are forwarded to the two APs in accordance to scheduling policy; some are forwarded to one AP, while the others are forwarded the other AP. However, they are not forwarded in the conventional way. Instead, when the scheduler is ready to forward a packet from $F_D$, it forwards a random linear combination, i.e.. network coding \cite{Dejan2018,Lnetworkcoding}, of all cohort $k$ in the queue. Before forwarding, the packet header is modified to indicate that it is from cohort $k$ and it is also tagged ``high priority''. It proceeds in this fashion either until all cohort $k$ packets have been forwarded once or until the interval expires. Cohort $k$ packets are not discarded from the queue immediately after forwarding; instead they are used for coded taking.

\item {\bf Coded Taking.} If additional time is available until $(k+1)T$, $G$ also transmits redundant packets to both APs in accordance with scheduling policy. 
Note that, this additional time is present when the sum of the bandwidth available
for $F_D$ from $G$ to the two APs exceeds the current TCP throughput for $F_D$,
a common occurrence when the bottleneck is at the mmWave links. 
The redundant packets are obtained by random linear combination of
cohort $k$ packets. Prior to forwarding, the packet header is modified to indicate that it is from cohort $k$. These packets are also tagged as ``low priority''. At time $(k+2)T$, all cohort $k$ packets are discarded. 
\end{itemize}

\begin{algorithm}[h]
\small
{\bf When a packet in $F_{D}$ arrives on incoming link $l$:}
\begin{algorithmic}[1]
\IF{buffer is full,}
\IF{incoming packet is marked as ``Low priority'',}
\STATE [Give: ]  Discard the incoming packet.
\ELSIF{all packets in buffer are marked as ``High priority'',}
\STATE [Give: ]    Discard the incoming packet.
\ELSE
\STATE [Give: ]    Discard the latest received ``Low priority'' packet in the buffer.
\STATE \newadd{Enqueue incoming packet.}
\ENDIF
\ELSE
\STATE \newadd{Enqueue incoming packet.}
\ENDIF
\STATE Let $\labelunsnt$ be the smallest unprocessed label of all $F_{D}$ packets in buffer.
\STATE Let $\labelrecv$ be the largest label of all $F_{D}$ packets received
            on incoming link $l$.
\IF{\newadd{$\labelunsnt \leq  \min \{\labelrecv : \mbox{incoming}~\link\} - 1$,}}
\STATE $\displaystyle \labelsnt = \labelunsnt$.
\STATE [Coding: ]  Discard all packets in $F_{D}$ in buffer with label smaller than $\labelsnt$.
\ENDIF

\hspace{-18pt}\noindent\textbf{\newadd{On the outgoing link of each AP:}}\\
\hspace{-12pt}\noindent Transmit (as permitted by packet scheduler)
          linearly \\
          \hspace{-12pt}\noindent combined packets in $F_{D}$ with label
          $\labelsnt$ in buffer no more \\ \hspace{-12pt}\noindent than number of received packets of label $\labelsnt$.
\IF{the total number of packets sent by all outgoing links is less than $n_{h}$,}
\STATE [Coding: ]  Duplicate and forward an unsent ``High priority'' packet with label $\labelsnt$.
\STATE [Give: ] Conditionally forward the unsent ``Low priority'' packet with label $\labelsnt$.
\ELSE
\STATE [Take: ]  Create a random linearly combined packet of packets in $F_{D}$ in buffer.
\STATE [Take: ]  Attach a label $k$, mark it as ``Low priority'', and forward it.
\ENDIF
\end{algorithmic}
\caption{Algorithm executed by the \textbf{mmWave Access Point} in \textbf{\method}.}
\label{alg:mnode}
\end{algorithm}


\noindent
{\textbf{Operations at} \bf Each AP:} Algorithm.~\ref{alg:mnode} describes the operations at each AP. Cohort $k$ packets are queued \newadd{and} not forwarded until the first cohort $(k+1)$ packet arrives.  Forwarding occurs with following mechanism.
\begin{itemize} 
\item {\bf Giving.} Forwarding of cohort $k$ packets occurs immediately after the arrival of the first cohort $k+1$ packet and {\em only until} the first cohort $k+2$
packet. As soon as the first cohort $k+2$ packet arrives, the remaining cohort $k$
packets in the queue are discarded, even if they were never forwarded. This important
aspect of the proposed schemed is called {\em Giving}, because unlike conventional
routers, it may give away packets without forwarding when bandwidth is not available. 

\item {\bf Coded Forwarding.} High priority cohort $k$ packets are forwarded to mobile device $D$ over the mmWave link in accordance with scheduling policy.

\item {\bf Coded Taking.} If additional time is available until first cohort $k+2$ packet
arrives, AP also forwards linear combination of cohort $k$ packets in the queue. 
\end{itemize}

\begin{algorithm}
\small
{\bf When a packet in $F_{D}$ arrives on incoming link $l$:}
\begin{algorithmic}[1]
\IF{buffer is full,}
\IF{incoming packet is marked as ``Low priority'',}
\STATE [Give: ]  Discard the incoming packet.
\ELSIF{all packets in buffer are marked as ``High priority'',}
\STATE [Give: ]    Discard the incoming packet.
\ELSE
\STATE [Give: ]    Discard the latest received ``Low priority'' packet in the buffer.
\STATE \newadd{Enqueue incoming packet.}
\ENDIF
\ELSE
\STATE \newadd{Enqueue incoming packet.}
\ENDIF
\STATE Let $\labelunsnt$ be the smallest unprocessed label of all $F_{D}$ packets in buffer.
\STATE Let $\labelrecv$ be the largest label of all packets in $F_{D}$ received on incoming link $l$.
\IF{\newadd{$\labelunsnt \leq  \min \{\labelrecv : \mbox{incoming}~\link\} -1$,}}        
\STATE $\displaystyle \labelsnt = \labelunsnt$.
\STATE [Coding: ]   Discard all packets in $F_{D}$ in buffer with label smaller than $\labelsnt$.
\STATE [Coding: ]   Apply decoding algorithm on packets in $F_{D}$ with label $\labelsnt$.
\IF{the decoding algorithm succeeds,}
\STATE [Coding: ]    Transmit the recovered unencoded
          packets on the outgoing links as per routing algorithm.
\ELSE
\STATE [Coding: ] Discard undecoded $F_{D}$ packets with label $\labelsnt$.
\ENDIF
\ENDIF
\end{algorithmic}
\caption{Algorithm executed by the \textbf{Mobile Device} in \textbf{\method}.}
\label{alg:access}
\end{algorithm}


\noindent
{\textbf{Operations at} \bf Mobile Device \bm{$D$}:} 
Algorithm.~\ref{alg:access} describes ``Algorithm Mobile Device'', the operations executed by the mobile device $D$.
Similar to $AP_i$, the mobile device $D$ also follows the label value of the network encoded packets of $F_D$ to determine which packets to decode. The mobile device $D$ decodes packets with label $k$ when each incoming link has received a packet in $F_{D}$ with label greater than $k$, the smallest unprocessed label in buffer. If packets are successfully decoded, the unencoded packets are forwarded to the application layer for immersive content streaming. Otherwise, undecoded packets are discarded.

\section{Evaluation}
\newlength\heightfiga\newlength\heightcapa
\newlength\heightfigb\newlength\heightcapb
\newlength\heightfigc\newlength\heightcapc
\newlength\heightfig

\label{sec:eval}
\ICC{
The performance of \method\ is evaluated using a network simulator. The network simulator uses measurements from a wearable VR setup for link bandwidth fluctuations. The setup measures the link bandwidth fluctuations caused by \ICCadd{changes in user poses, locations, and directions}
while participating in an immersive experience on an Oculus Quest 2 headset.  Although the Oculus Quest 2 headset does not support IEEE 802.11ay mmWave wireless link, the wearable setup includes 802.11ay devices deployed to capture the effects of user movements on the mmWave link bandwidth.}
\ICC{The simulations
explore the impact of user movements 
on transport layer performance \ICCadd{during the immersive VR experience}.
We compare the performances of MuSher and \method\ for different \ICCadd{network delays} and packet error rates (PER) in the network simulator. }

\subsection{\ICCadd{Experimental Setup}}
\textit{Wearable real-time VR mmWave channel measurement system:}
 We first build the wearable mmWave link system to measure the mmWave channel capacity at real time during the use of VR headset.
\subsubsection{\ICCadd{VR scenario setup}}
\begin{figure}[!h]
\centering
\begin{subfigure}[b]{0.49\columnwidth}
  \centering
  \includegraphics[width=\columnwidth]{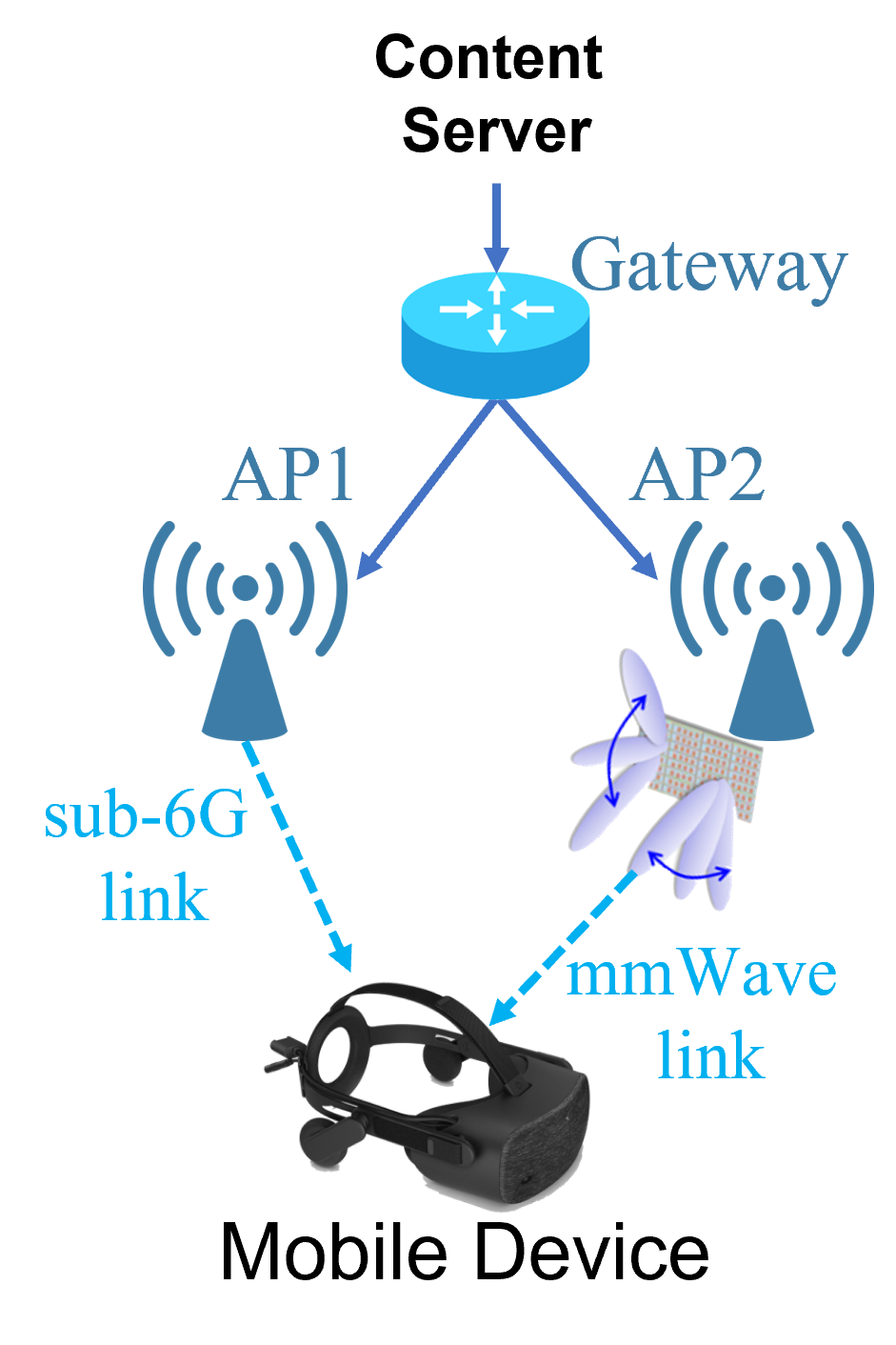}  
  \caption{\label{fig:SetupTopo} Dual connectivity topology}
\end{subfigure}
\begin{subfigure}[b]{0.49\columnwidth}
  \centering
  \includegraphics[width=\columnwidth]{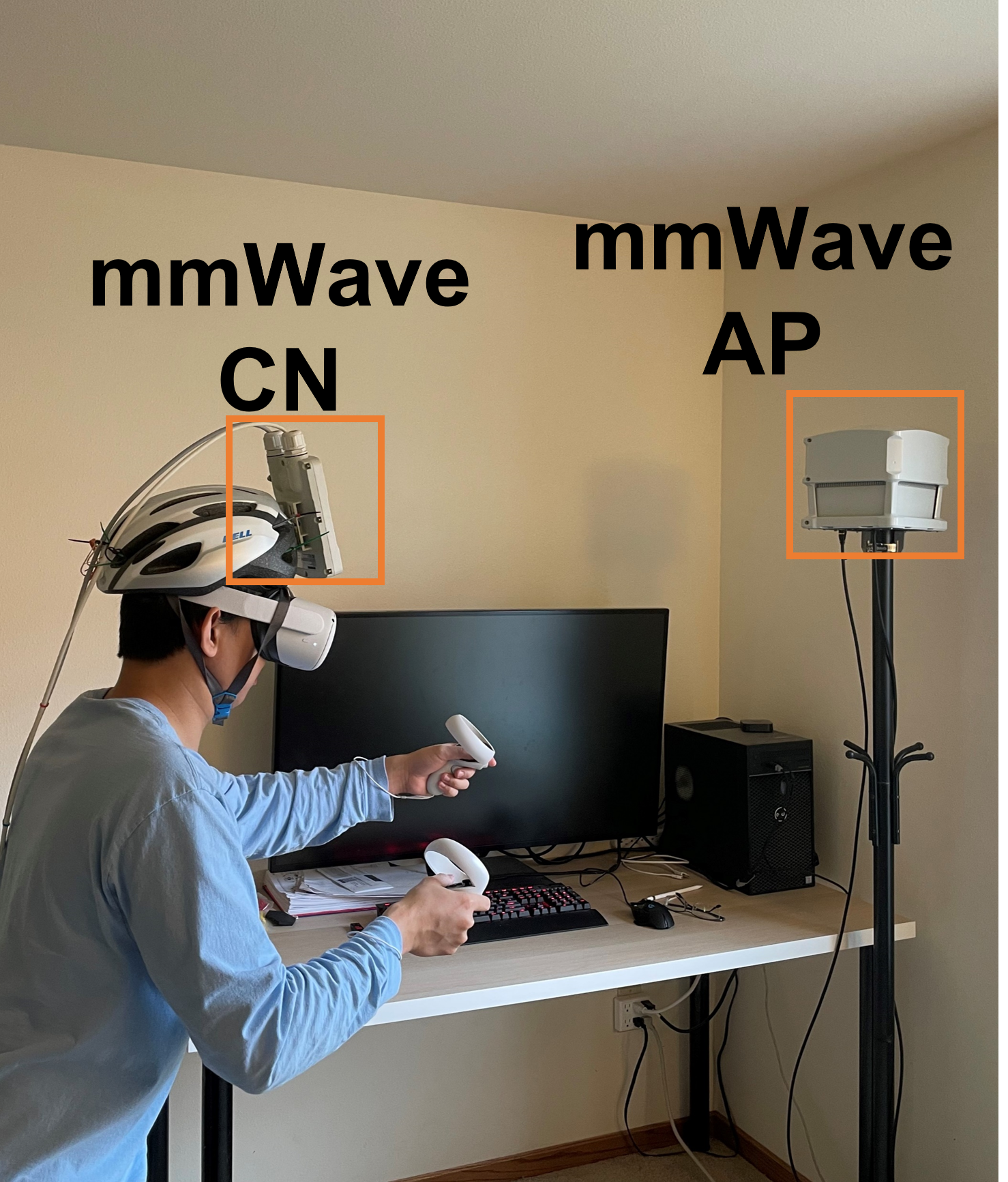}  
  \caption{\label{fig:setup} Augmented VR setup}
\end{subfigure}
\caption{\label{fig:TotalSetup} mmWave channel measurement setup}
\vspace{-15pt}
\end{figure}

The dual connectivity scenario of mmWave network simulations are shown in Fig.~\ref{fig:SetupTopo}. 
This dual connectivity involves one sub-6G connection and one mmWave connection with modern beamforming techniques. These two connections are supported by two different APs, which are connected to the same gateway. The mobile device associates with these two connections to establish dual link connectivity and receives packets from both mmWave and sub-6G APs simultaneously. \ICC{Since the user movements are fairly small, the sub-6G connection seems to provide a relatively stable channel bandwidth.}

\ICCadd{Fig.~\ref{fig:setup} shows a picture of a user in an immersive VR experience.
The setup augments Oculus Quest 2 headset with additional mmWave devices because the current headset does not support mmWave connections. The mmWave link channel quality fluctuations are obtained from measurements while the user is in the immersive experience. The user wears both the VR headset and the mmWave client node (CN) on the head and the mmWave channel capacity between the AP and the CN is measured in real time while the user is using VR applications.}
To measure the \ICCfinal{network-layer} link bandwidth fluctuations in real-time which the user is participating in a VR experience, we use a well-known technique called packet trains \cite{pktTrain}.

\begin{figure}
    \vspace{-15pt}
    \centering
    \includegraphics[width=\linewidth]{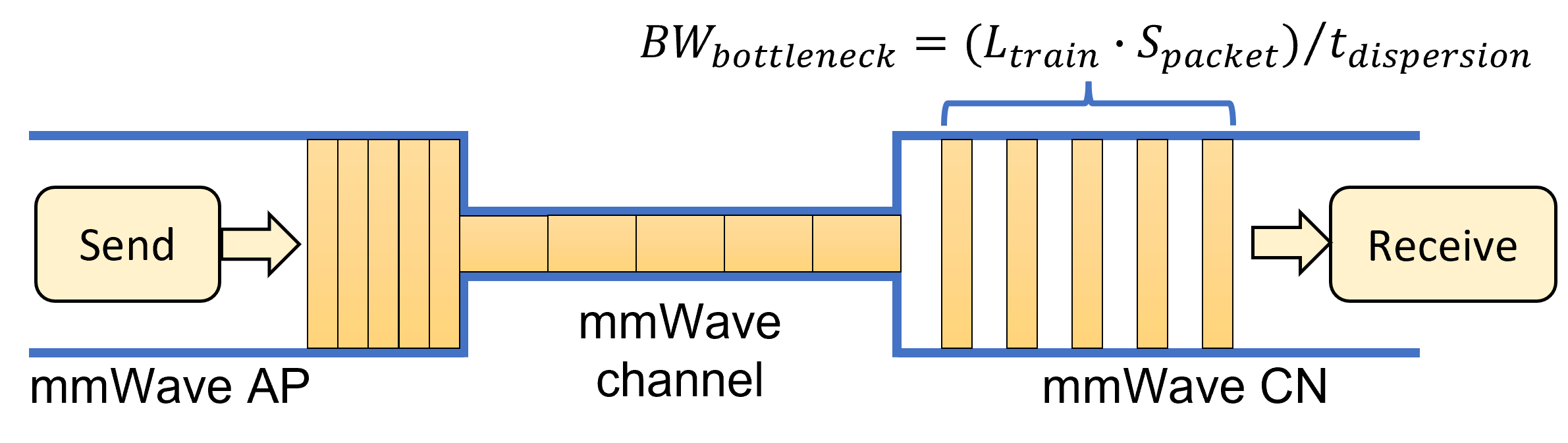}
    \caption{Packet train illustration}
    \label{fig:PktTrain}
\end{figure}

\ICC{
\subsubsection{Packet train based measurement of link bandwidths}
Packet train is an extension of a concept called packet pair \cite{pktTrain}. A packet pair measurement system is comprised of a sender and a receiver. The network route from the sender to the receiver goes through the link whose bandwidth is to be measured. The packet pair technique assumes that desired link has the smallest bandwidth among all the links in the route. The packet pair sender transmits two back-to-back measurement packets. The technique relies on the observation that often these two packets get queued back-to-back waiting for transmission on the desired link and then get transmitted back-to-back on the desired link.  When this happens, the spacing between the two packets (called the time dispersion) is inversely proportional to the link bandwidth. Due to different measurement errors, the link bandwidth may not be correctly estimated one packet pair.  However, a good estimate can be obtained from repeated packet pair measurements. 
When the link bandwidth is high, as in the case of mmWave links, the time dispersion is very small and the receiver clock may not be accurate enough to measure time dispersion accurately. 

To overcome this problem, one can use a train of measurement packets, instead of a pair.  This ideas is illustrated in Fig.~\ref{fig:PktTrain}.}
One packet train is a group of UDP packets having the same size. 
\ICC{As shown in the figure, ideally, all the packets in the train are queued and then later transmitted back-to-back on the mmWave link.}
Once the \ICC{receiver} receives a packet train, it calculates the bottleneck bandwidth along the path according to the train length and the time dispersion in the bottleneck bandwidth. If $L_{train}$ is the length of one packet train and the UDP packet size is $S_{packet}$, \ICC{then relationship to link bandwidth can be shown to be $BW_{bottleneck} = (L_{train} \cdot S_{packet})/t_{dispersion}$.}
\ICC{The errors in measurement depend on the length of the packet train and the length of each packet in the train. Long packet trains and long packets tend to have smaller errors. However, they also increase the overhead of measurement. Through targeted simulations where one could quantify the errors, we chose  the following paramters: $L_{train}$ = 30 packets, $S_{packet}$ = 1500 Bytes. Since the link bandwidth fluctuates, we chose to repeat the packet train measurement once every 10 ms.}
\ICC{Note that, overhead of this measurement is only 36 Mbps for an approximately 1 Gbps link.} 
The trace of channel measurements from this setup was input to simulations in the form of data rates. 

\subsection{Performance Evaluation}

\response{
\ICCadd{We set different network delays and packet error rates (PER) and measure} the short-term TCP throughput and latency of MuSher and \method\ during the burst of immersive data transmitting in real mmWave channels. In simulations, TCP-CUBIC \cite{Ha2008} is used to assure reliable, sequenced delivery of packets with the packet size of 128 bytes. The channel bandwidth \ICCfinal{observed at the network layer} is set in the form of data rate value and the bandwidth changes according to the packet train measurement interval, which is 10 ms\ICCfinal{, reflecting the network-layer channel quality with the impacts at MAC and physical layers}. Results are shown in Fig.~\ref{ShortTermCompare} and \ref{Time}.



\begin{figure}[!h]
\centering
\begin{subfigure}{0.49\columnwidth}
  \centering
  \includegraphics[width=\columnwidth]{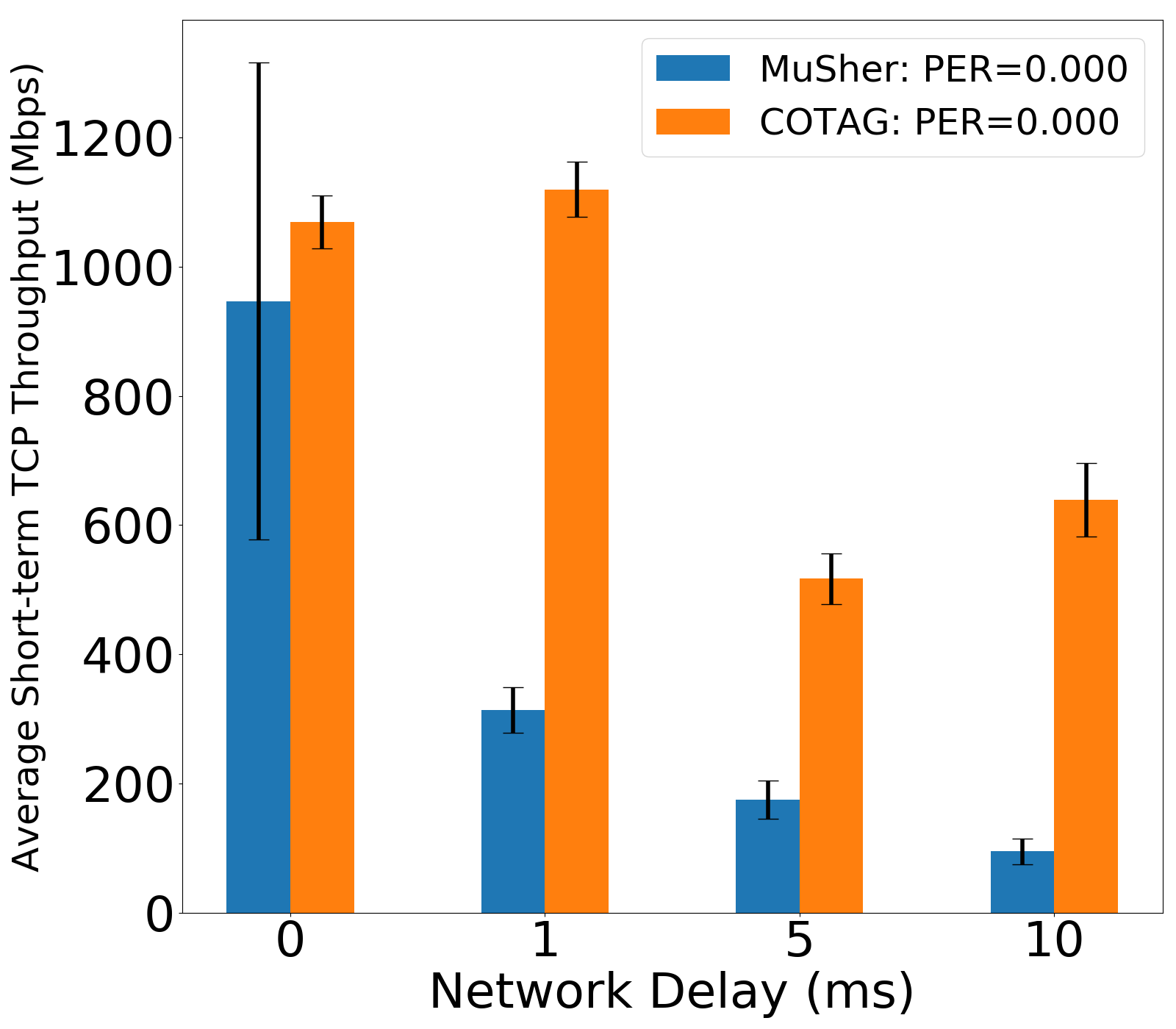}  
  \caption{\label{SeptopoCompare} Results with zero PER}
\end{subfigure}
\begin{subfigure}{0.49\columnwidth}
  \centering
  \includegraphics[width=\columnwidth]{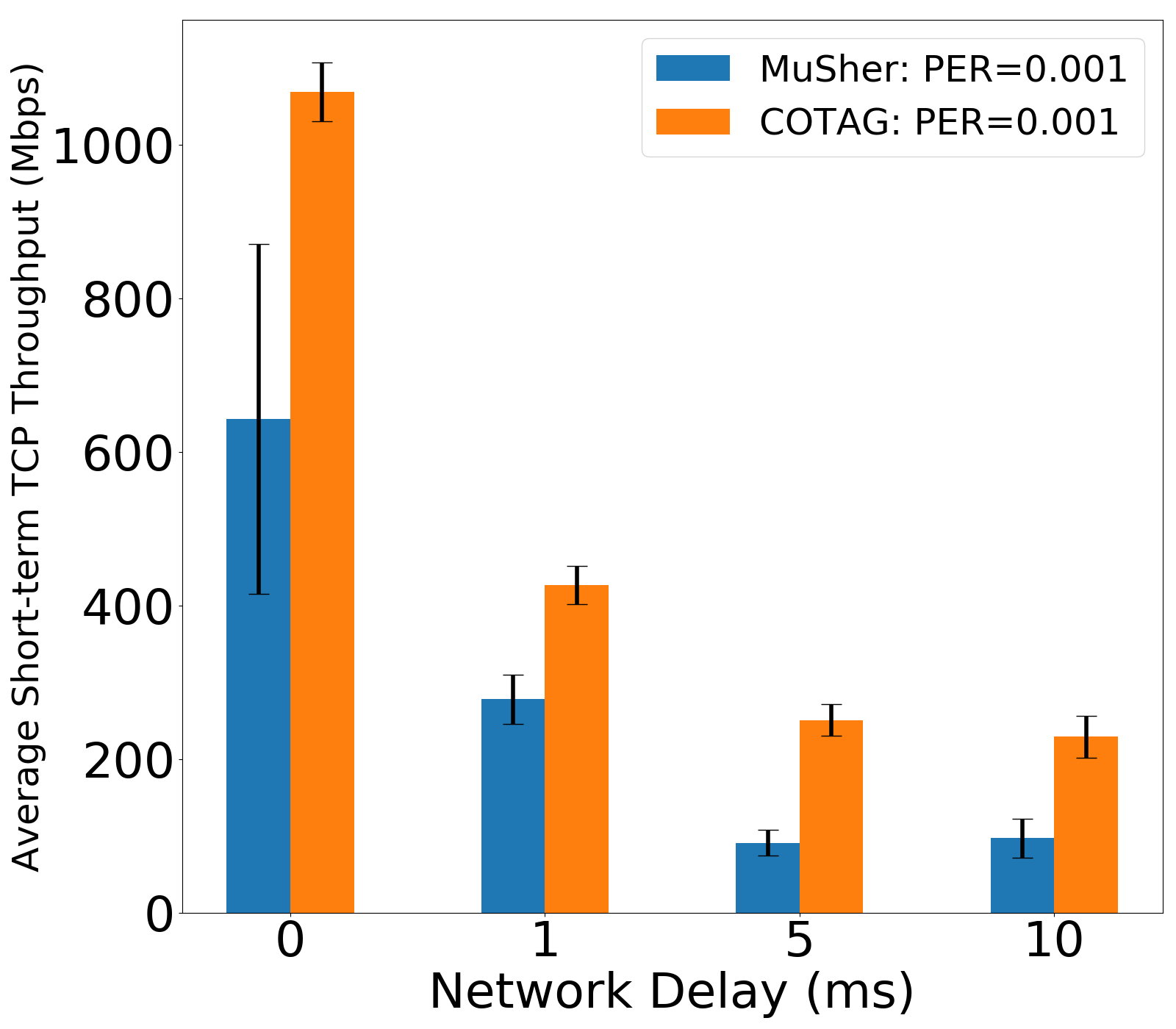}  
  \caption{\label{EqtopoCompare} Results with 0.001\% PER}
\end{subfigure}
\caption{\label{ShortTermCompare} Short-term throughput with different PER}
\vspace{-15pt}
\end{figure}

\response{
Results in Fig.~\ref{SeptopoCompare} are based on simulations of MuSher and \method\ with zero PER in the dual connectivity network and with different \ICCadd{network delays in the back end} network. The error bars in the results show the range of 90\% confidence intervals. In this case, the impacts are the fluctuations of mmWave channel bandwidth, the bandwidth unbalance between the mmWave channel and sub-6G channel and the \ICCadd{delay in the back end network}. It shows that as the \ICCadd{delay} in the network grows, the impact of the channel fluctuations and bandwidth unbalance degrades TCP performance more severely. 
Simulations in Fig.~\ref{EqtopoCompare} shows the average short-term throughput of MuSher and \method\ with 0.001\% PER in wireless channels. The results show that with additional impact from PER, TCP performance is more vulnerable as the \ICCadd{network delay} increases.
These results in Fig.~\ref{ShortTermCompare} show that in real VR application scenario, \method\ significantly enhances TCP's short-term throughput.


\begin{figure}[!h]
\centering
\begin{subfigure}{0.49\columnwidth}
  \centering
  \includegraphics[width=\columnwidth]{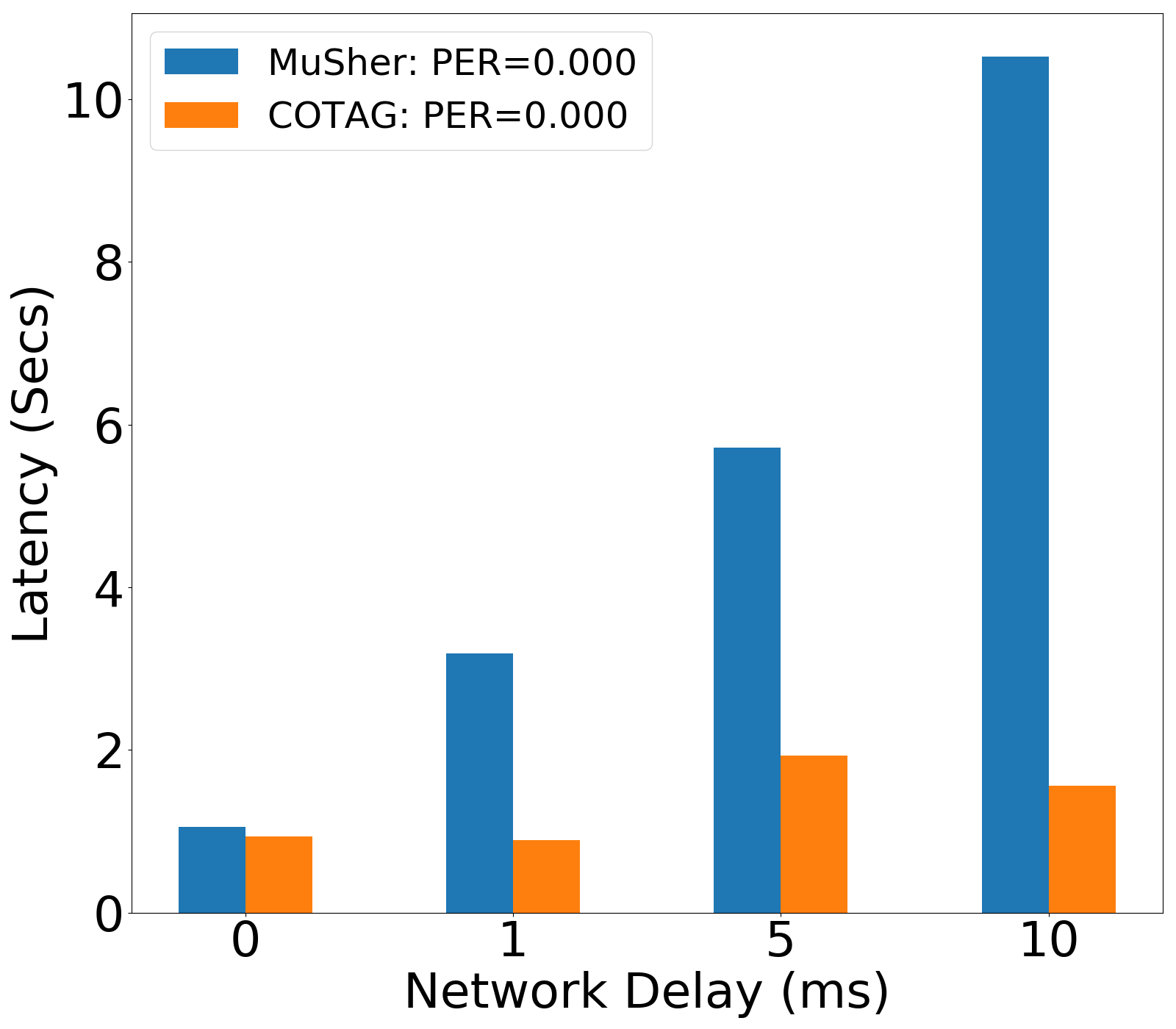}  
  \caption{\label{Time0} Results with zero PER}
\end{subfigure}
\begin{subfigure}{0.49\columnwidth}
  \centering
  \includegraphics[width=\columnwidth]{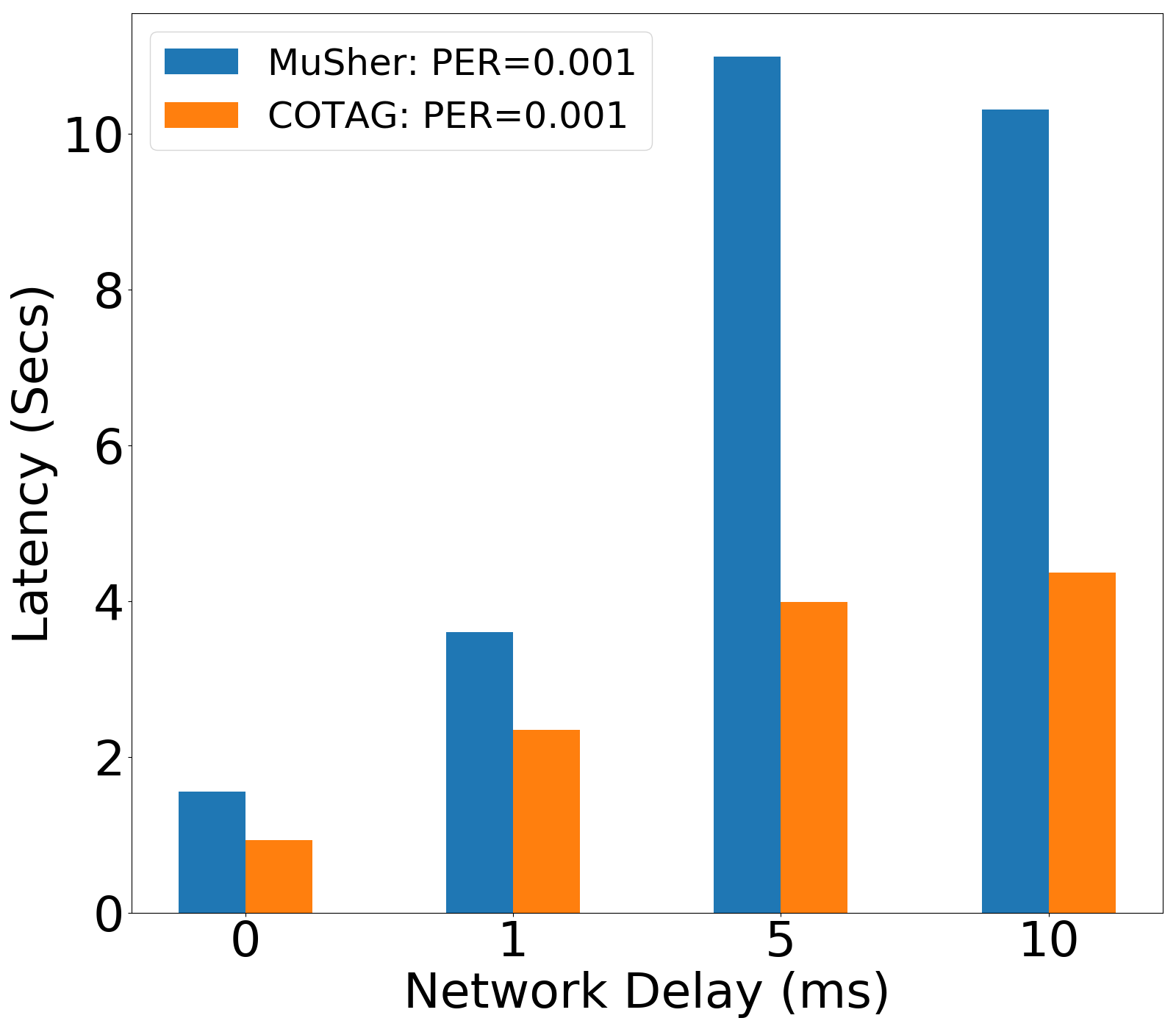}  
  \caption{\label{Time1} Results with 0.001\% PER}
\end{subfigure}
\caption{\label{Time} Data burst delivery latency with different PER}
\vspace{-8pt}
\end{figure}

\ICCadd{Moreover, the user movements and rotations can cause changes in the VR headset viewpoint during the use of immersive VR applications. In this case, a burst of data needs to be delivered for the new viewpoint. If the data is not delivered quickly, the user experience may be degraded. Thus, we evaluate the latency in delivering a burst of data.  With the same PER and network delay setting, Fig.~\ref{Time} presents the average latency for transmitting 1 Gb sequenced data when there is a data burst in the immersive application. In these simulations, \method\ has much smaller latency that MuSher.}

}
}

\section{Conclusion}

\label{sec:summary}
Dual connectivity transmission based on mmWave and sub-6G wireless links is considered for delivering immersive content to AR/VR applications. 
To better enhance the robustness of dual connectivity transmission on network layer, we propose a \fullmethod\ (\method) scheme in the presence of user movements, rotations and blockages. 
The simulation results show that considering network latency, packet errors, and link quality unbalance and fluctuations in dual connectivity, when traditional transport layer solutions are affected, 
\method\ can \ICCadd{enhance TCP's short-term throughput and latency for short data bursts.}

\section*{Acknowledgments}
This work was partially supported by National Science Foundation grant CNS 1703389, U.S.A and by grant MOST 108-2221-E-011-058-MY3 of the Ministry of Science and Technology, Taiwan, R.O.C.
\small
\bibliographystyle{IEEEtran}
\bibliography{references}

\end{document}